# Gatekeeping: a Partial History of Cold Fusion

Jonah F. Messinger, Florian Metzler and Huw Price

**Abstract:** One of the most public episodes of gatekeeping in modern science was the case of so-called 'cold fusion'. At a news conference in 1989 the electrochemists Martin Fleischmann and Stanley Pons announced that they had found evidence of nuclear fusion in palladium electrodes loaded with deuterium. There was worldwide interest. Many groups sought to reproduce the results, most unsuccessfully. Within months, the prevailing view became strongly negative. The claims of Fleischmann and Pons came to be regarded as disreputable, as well as false. As the Caltech physicist David Goldstein put it, cold fusion became 'a pariah field, cast out by the scientific establishment' (Goldstein 1994). The case would already be interesting for students of gatekeeping if the story had ended at that point. Even more interestingly, however, the field survived and persisted. It has been enjoying a modest renaissance, with recent government funding both in the US and the EU. This piece offers an opinionated introduction to cold fusion as a case study of scientific gatekeeping, discussing both its early and recent history.

## 1. Lessons from controversial cases

Students of gatekeeping need good case studies. We can theorize about how science should police its borders, but what happens in practice? This piece is an opinionated introduction to one such example – a famous one, in some respects.

Most readers will know something of the history and reputation of so-called 'cold fusion'. Some will recall its spectacular rise and fall in 1989. Many will be surprised to learn that it is not a closed case (far from it). The current state of the field makes it especially interesting for students of gatekeeping, in our view; but there are important lessons in its early history, even without the benefits of hindsight.

Here, as in general, the study of gatekeeping is potentially a controversial business. If we have the normative question in mind – if we want to ask whether gatekeepers got particular cases *right* – then we put on the table the possibility of a negative answer. We may then attract the attention of the border police ourselves, at least in cases of current interest.

Cold fusion offers some of this metascientific risk. Some readers may think us foolish for being open-minded about a case they regard as not merely closed, but rightly derided. If so, then this reaction is a manifestation of phenomena we want to discuss. The case in question has certainly been controversial, but that's part of what makes it interesting, for students of gatekeeping. If we can confirm this by making the hair of some of our readers stand on end, so much the better.

The rest of this paper goes like this. The next section (§2) offers a very selective history of the field in its first five years (1989–1994), focusing on aspects we take to be relevant to gatekeeping. §3 and §4 then draw some critical conclusions. In some respects, we argue, the field provides remarkable examples of how to get gatekeeping wrong. As promised, these criticisms don't depend on hindsight, though they are reinforced by later developments. Moving towards the present, §5 deals briefly with the twenty years 1995–2014, flagging a few episodes relevant to our main argument, and §6 describes a very interesting decade since 2015. §7 identifies some possible future paths for the field, and §8 draws some conclusions.

In sketching the early history, we'll make use of a (1994) essay by a Caltech physicist, David Goodstein (1939–2024). Goodstein's title – 'Pariah Science: Whatever Happened to Cold Fusion?' – conveys the reputation that the subject had acquired at that point. But Goodstein's perspective was very unusual. He had friends and colleagues on both sides of what soon became a very high fence, between cold fusion and mainstream science. He was also Vice Provost at Caltech, with responsibility for investigations of allegations of scientific fraud.

Goodstein was also an accomplished science writer and educator. His (1994) essay became the basis for a chapter in his (2010) book, *On Fact and Fraud: Cautionary Tales from the Front Lines of Science.* The chapter is titled 'The Cold Fusion Chronicles'. Goodstein doesn't use the term 'gatekeeping', but his closing paragraph is an affirmation of the lasting interest of cold fusion, whatever its future, as an example for students of scientific practice.

> [T]he cold fusion saga offers a classic case study of how scientists … may convince themselves that they are in the possession of knowledge that does not in fact exist.[1] This is not scientific misconduct, but it is a phenomenon of considerable scientific and human interest. Many mistakes were made on both sides … of the cold fusion story, but they were honest errors, not … fraud. I suppose that if nuclear fusion really has taken place in some of these experiments, the world of conventional science will eventually be forced to take notice. If not, then the whole episode will remain a curious and instructive footnote in the history of science. Either way, it illuminates the inner dynamics of the scientific enterprise in a way that few other stories have done. (Goodstein 2010, 95)

On present indications, as we will show below, history is taking the first of these paths. If so, then early scientific work tagged with the label 'cold fusion' may come to be seen as the stumbling beginnings of a *new* conventional science, one that deals with the impact of condensed matter environments on nuclear processes and reactions.

Some of the present authors work within this new science, as we take it to be. This means that in writing about gatekeeping, our perspective is neither detached nor purely historical. The history is still in process, and we ourselves not only have views about its direction, but hope to contribute to it. What we offer is a *partial* history, in both senses of the word – an early, internal, history-in-progress of the new field of condensed matter nuclear science.[2]

---

[1] Goodstein may have intended this as a one-sided comment, meaning that it is the proponents of cold fusion who convinced themselves of 'knowledge that does not in fact exist'. But it also works in a more even-handed spirit, allowing for the possibility that it was the gatekeepers' confidence that was misplaced. That fits better with the remainder of the paragraph, and we'll give Goodstein the benefit of the doubt here, reading him in the even-handed way.

[2] While 'condensed matter nuclear science' (in use since 2002) is perhaps the most descriptive and least normative term for the field under discussion here, we will continue to use the familiar term 'cold fusion' in this piece, without meaning to imply that it would be an accurate description of the core phenomena of such a science. We'll also use 'LENR', for 'low energy nuclear reactions', another term that has come into common use.

## 2. The first five years (1989–1994)

### 2.1 Goodstein on pariah science

Goodstein's (1994) essay begins with a description of the Fourth International Conference on Cold Fusion, held in Maui, Hawaii, in December 1993. Goodstein notes that the event 'had all the trappings' of an ordinary scientific meeting. But, contrary to appearances, he continues, 'this was no normal scientific conference':

> Cold Fusion is a pariah field, cast out by the scientific establishment. Between Cold Fusion and respectable science there is virtually no communication at all. Cold Fusion papers are almost never published in refereed scientific journals, with the result that those works don't receive the normal critical scrutiny that science requires. On the other hand, because the Cold Fusioners see themselves as a community under siege, there is little internal criticism. Experiments and theories tend to be accepted at face value, for fear of providing even more fuel for external critics, if anyone outside the group is bothering to listen. In these circumstances, crackpots flourish, making matters worse for those who believe that there is serious science going on here. (1994, 527)

There are already gatekeeping lessons here, to which we'll return. This is Goodstein's description of the origins of the field, in March 1989:

> The origins of Cold Fusion have been loudly and widely documented in the press and popular literature. [Stanley] Pons and [Martin] Fleischmann, fearing they were about to be scooped by a competitor named Steven Jones from nearby Brigham Young University, and with the encouragement of their own administration, held a press conference on March 23, 1989, at the University of Utah, to announce what seemed to be the scientific discovery of the century. Nuclear fusion, producing usable amounts of heat, could be induced to take place on a tabletop by electrolyzing heavy water using electrodes made of palladium and platinum, two precious metals. If so, the world's energy problems were at an end – to say nothing of the fiscal difficulties of the University of Utah. What followed was a kind of feeding frenzy, science by press conference and E-mail, confirmation and disconfirmations, claims and retractions, ugly charges and obfuscation, science gone berserk. (1994, 528)

'For all practical purposes', Goodstein continues, this feeding frenzy didn't last very long.

> [I]t ended a mere five weeks after it began, on May 1, 1989, at a dramatic session of The American Physical Society, in Baltimore. Although there were numerous presentations at this session, only two truly counted. Steven Koon and Nathan Lewis, speaking for himself and Charles Barnes, all three from Caltech, executed between them a perfect blocked shot that cast Cold Fusion right out of the arena of mainstream science. (1994, 528)

This assessment is a little Caltech-centric, in our view. As we'll see, there was mainstream attention to cold fusion, albeit much of it negative, for many months after the beginning of May.[3]

Goodstein's narrative now takes its unusual turn. As he says, he counted among his friends not only these Caltech colleagues, but also an Italian physicist, Francesco ('Franco') Scaramuzzi. Goodstein describes the experimental work of Scaramuzzi and his colleagues, initially involving the detection of neutron emissions from devices based on those of Pons and Fleischmann, and later, in 1992–1993, the detection of excess heat.

Goodstein recalls that he visited Scaramuzzi in Rome in December 1993, and heard him present his work to sceptical colleagues at the University of Rome:

> I went to visit my friend Franco in December 1993, when he returned from the Maui conference. While I was there, he summarized the results of the conference in a seminar presented to the physics faculty at the University of Rome … . This was in itself an unusual event. The physics faculty of the University of Rome today is comparable to the physics department at a good American state university. For them, inviting Franco to speak about Cold Fusion was a daring excursion to the fringes of science. Feeling this was a rare opportunity, Franco prepared his talk with meticulous care.
>
> At the seminar, Franco's demeanor was subdued, and his presentation was, as always, reserved and correct. Nevertheless, his message was an optimistic one for Cold Fusion. In essence …, each of the criticisms that Nathan Lewis had correctly leveled at the experiments of Pons and Fleischmann had been successfully countered by new experiments reported at the conference. Even more important, there was reason to believe that the magic missing factor, the secret ingredient of the recipe that accounted for why Cold Fusion experiments only sporadically gave positive results, might finally have been discovered. (1994, 538)

Goodstein describes three then-familiar criticisms from Lewis: first, that the calorimetry might have been affected by local hot spots in the electrolysis experiments (a familiar occurrence); second, that the observed heat might result from some chemical storage effect; and third, lack of sufficient control experiments. As Goodstein says, Scaramuzzi reported that all of these objections had been 'successfully countered'. And there was more:

> All of this was much less important than the fact that Cold Fusion experiments, if they gave positive results at all, gave them only sporadically and unpredictably. … The essential key to the return of Cold Fusion to scientific respectability is to find the missing ingredient that would make recipe work every time. (1994, 539–540)

Scaramuzzi reported new work suggesting that a critical factor was the ratio of deuterium atoms to palladium atoms in the samples.

---

[3] The Caltech perspective may also be reflected in Goodstein's apparently uncritical acceptance that the case could *possibly* be settled so quickly. This is surprising in the light of his own connections to alternative viewpoints, and regrettable for the reasons we explore in §4.3 below.

> Both the American and Japanese groups showed data indicating there is a sharp threshold at [the ratio] $x = 0.85$. Below that value (which can only be reached with great difficulty and under favorable circumstances) excess heat is never observed. But, once $x$ gets above that value, excess heat is essentially always observed, according to the reports … recounted by Franco Scaramuzzi … (1994, 540)

How was this news received?

> The audience at Rome, certainly the senior professors who were present, listened politely, but they did not hear what Franco was saying – that much became clear from the questions that were asked at the end of the seminar and comments that were made afterward. If they went away with any lasting impression at all, it was just the sad realization that so fine a scientist as Franco Scaramuzzi had not yet given up his obsession with Cold Fusion. They cannot be blamed. Any other audience of mainstream scientists would have reacted exactly the same way. If Cold Fusion ever regains the scientific respectability that was squandered in March and April of 1989, it will be the result of a long, difficult battle that has barely begun. (1994, 540)

Was Goodstein expressing regret about the early history of the cold fusion case?  In his later piece he says that 'mistakes were made on both sides of the great divide in the course of the cold fusion story, but they were honest errors, not examples of fraud.' (Goodstein 2010, 95) Could such mistakes have been avoided, if some metascientific lessons had been learnt in advance? Goodstein doesn't raise the question. What he offers to students of gatekeeping is a vivid description of the costs to a would-be scientific community if the borders are closed against them too harshly, and too soon.

## 2.2 When did mainstream science abandon cold fusion?

As noted, Goodstein's view is that the borders were closed to cold fusion by his Caltech colleagues on May 1, 1989, when they 'cast Cold Fusion right out of the arena of mainstream science'. But there is plenty of evidence that mainstream science was still paying attention, much but not all of it critical, for many months after that May Day in Baltimore.

In mid-October 1989, for example, there was a closed meeting in Washington, DC, co-sponsored by the Electric Power Research Institute (EPRI) and the National Science Foundation (NSF). This was a follow-up to a meeting in Santa Fe, New Mexico, in May 1989. As the Proceedings say, '[t]his second workshop … brought together skeptics and advocates to facilitate communication, to examine closely the experimental results, and to identify research issues.' (EPRI-NSF 1993) Keynote speakers included Edward Teller, and Goodstein's Caltech colleague, Nathan Lewis. Lewis appears to have made an effort not to be strongly adversarial:

> I would like to express my thanks to the organizers of the workshop for allowing me to describe my viewpoint concerning the electrochemistry of the palladium-$D_2O$ system. My task is to describe it without taking a stand for or against the alleged phenomena which we are here to discuss. … I hope to demonstrate what has been measured, what has not been measured, what controls are needed, and what other issues exist, particularly in respect to the electrochemical charging of palladium cathodes with deuterium, to measuring the heat flux, and to measuring separation factors for the isotopes of hydrogen.

    A substantial amount of hearsay and rumors have been heard concerning what might occur at this meeting and also concerning the citing of positive confirming results based on the work of laboratories whose results have not been discussed in scientific meetings or written up in peer-reviewed journals. I believe that we must avoid any of these pitfalls. Since I must be objective, I will discuss only those things which I feel that I definitely know, and which will therefore be from my own direct experience in my laboratory. I hope that the same philosophy will be adopted in other contributions to this workshop. (EPRI-NSF 1993, 2-1)

As for Edward Teller, his opening remarks are perceptive and open-minded, arguing for 'a continuation of an effort', but 'primarily for the sake of pure science', not based on the promise of immediate practical applications. Teller regards the latter question as premature.[4]

> I … applaud the National Science Foundation and the Electric Power Research Institute for maintaining enough interest and enough support so that a real clarification of the apparent contradictions can be pursued. If that clarification would lead to something on which we can agree and to a reaction probability which is small, but much bigger than the Gamow factor would allow, this would be a great discovery. Perhaps a neutral particle of small mass and marginal stability is catalyzing the reaction.
>
>     You will have not modified any strong nuclear reactions, but you may have opened up an interesting new field (i.e., the very improbable actions of nuclei on each other). So, I am arguing for a continuation of an effort, primarily for the sake of pure science. And, of course, where there is pure science, sometimes, at an unknown point, applications may also follow.
>
>     But, according to my hunch, this is a very unclear and low probability road into a thoroughly new area. The low probability has to be balanced against the great novelty. But to think beyond that and ask what is the practical application, what this very unknown area of nuclear physics may produce, that I claim, is completely premature. (EPRI-NSF 1993, 1-1–1-2)

There were some positive news reports about this meeting. A front-page piece in *The New York Times* is headed 'Recent Tests Said to Justify More Cold Fusion Research' (Leary 1989). It quotes the Co-Chairs of the meeting, Dr Paul Chu (University of Houston) and Dr John Appleby (Texas A&M University), saying: 'Based on the information that we have, these effects cannot be explained as a result of artifacts, equipment error or human errors.'

    In the popular perception, however, this EPRI-NSF meeting seems to have been eclipsed by strongly negative coverage elsewhere. There had been an on-going investigation by

---

[4] In the same remarks, Teller notes that 'the history of science and experimental physics is full of examples of predictions that things are impossible and yet they have happened' and he recalls Ernest Lawrence's joke: 'When Teller says it is impossible, he is frequently wrong. When he says it can be done, he is always right.' (EPRI–NSF 1993, 1-1–1-2)

the US Department of Energy (DoE), which released its report, largely negative, in early November 1989. (A preliminary version of the report had been released in July, as some of the news coverage of the October EPRI-NSF meeting notes.) And there was continuing negative coverage in the editorial pages of *Nature*.

In our view, both the DoE report and the *Nature* coverage display a surprising blindness to relevant issues of epistemic risk – clearly a factor relevant to gatekeeping, by contemporary standards. We will explain this point in the next section, with reference to the work of the philosopher Heather Douglas (Douglas 2000, 2007, 2009).

In the section after that (§4), setting issues of risk aside, we want to comment on the speed with which mainstream science formed its judgement about cold fusion. Even if Goodstein gives his Caltech colleagues too much credit, in claiming that they closed the door on the topic in early May, it still happened very rapidly. Was this appropriate, given the subject matter? Were there alternatives? As we'll see, the role of claimed failures of 'replication' is especially interesting for students of gatekeeping.

## 3. Big bang, low bar

### 3.1 Douglas on inductive risk

A familiar prudential principle tells us that the more harmful the consequences of some possible risk, the more improbable it needs to be, before we can reasonably set it aside. The following formulation, stressing the sense of responsibility, is from Heather Douglas.

> In general, if there is widely recognized uncertainty and thus a significant chance of error, we hold people responsible for considering the consequences of error as part of their decision-making process. Although the error rates may be the same in two contexts, if the consequences of error are serious in one case and trivial in the other, we expect decisions to be different. Thus the emergency room avoids as much as possible any false negatives with respect to potential heart attack victims, accepting a very high rate of false positives in the process. … In contrast, the justice system attempts to avoid false positives, accepting some rate of false negatives in the process. Even in less institutional settings, we expect people to consider the consequences of error, hence the existence of reckless endangerment or reckless driving charges. (Douglas 2007, 124)

Douglas goes on to discuss the possibility that '[w]e might decide to isolate scientists from having to think about the consequences of their errors', but rejects it. She argues that 'we want to hold scientists to the same standards as everyone else', and therefore 'that scientists should think about the potential consequences of error.'

Clearly, this principle is relevant to gatekeeping. The greater the possible harmful consequences of mistaken *exclusion* of a new body of claims, the lower the appropriate bar for their continued admission – 'Big Bang, Low Bar', as (Price 2024b) puts it. And harmful consequences can be indirect – a matter of missing something good, rather than encountering something bad. So missing out on potentially new ways to access abundant energy cleanly and cheaply might well qualify. Let's keep this in mind, as we revisit two of the powerful negative voices against cold fusion in 1989–1990.

3.2 The DoE report

The DoE report was commissioned by the US Secretary of Energy, in April 1989.

> In recent weeks, there has been a great deal of interest in the prospects for "cold fusion", based on experiments at universities in Utah and subsequent experiments performed elsewhere. … Because of the potential benefits from practical fusion energy, I request that the Energy Research Advisory Board (ERAB) assess this new area of research. (Letter from the US Secretary of Energy to the Chairman of the Energy Research Advisory Board, US Department of Energy, 24/4/89; DoE 1989, 39)

The resulting Panel delivered its final report on November 8, 1989. The report describes the Panel's work program as follows.

> The Panel or subgroups thereof have participated in the [first EPRI-NSF] Workshop on Cold Fusion in Santa Fe, have visited several laboratories, have examined many published articles and preprints, studied numerous communications and privately distributed reports, and have participated in many discussions. In addition, the Panel held five public meetings where its findings were discussed and drafts of both the Interim and Final Reports were formulated. (DoE 1989, 1)

Turning to its conclusions, the Panel says this:

> [W]ith the many contradictory existing claims it is not possible at this time to state categorically that all the claims for cold fusion have been convincingly either proved or disproved. Nonetheless, on balance, the Panel has reached the following conclusions and recommendations. (DoE 1989, 2)

We will reproduce here two of the conclusions and three of the recommendations, using the Report's own numbering.

> CONCLUSIONS
> (1) Based on the examination of published reports, reprints, numerous communications to the Panel and several site visits, the Panel concludes that the experimental results of excess heat from calorimetric cells reported to date do not present convincing evidence that useful sources of energy will result from the phenomena attributed to cold fusion.
> …
> (5) Nuclear fusion at room temperature, of the type discussed in this report, would be contrary to all understanding gained of nuclear reactions in the last half century; it would require the invention of an entirely new nuclear process.

RECOMMENDATIONS

(1) The Panel recommends against any special funding for the investigation of phenomena attributed to cold fusion. Hence, we recommend against the establishment of special programs or research centers to develop cold fusion.

(2) The Panel is sympathetic toward modest support for carefully focused and cooperative experiments within the present funding system.

(3) The Panel recommends that the cold fusion research efforts in the area of heat production focus primarily on confirming or disproving reports of excess heat. Emphasis should be placed on calorimetry with closed systems and total gas recombination, use of alternative calorimetric methods, use of reasonably well characterized materials, exchange of materials between groups, and careful estimation of systematic and random errors. Cooperative experiments are encouraged to resolve some of the claims and counterclaims in calorimetry. (DoE 1989, 2–3)

Some comments. First, these conclusions are not strongly positive, but not entirely dismissive, either. There is nothing in the report itself to support the claim that cold fusion had been 'cast out of the arena of mainstream science', as Goodstein put it, even if some of the Panel members may have hoped to achieve that.[5] Second, conclusion (1) amounts to 'no convincing evidence for a yes', not 'convincing evidence for a no'. Yet it is the latter that matters, if we regard the case as one with a very high potential cost of a false negative.

Compare Douglas's courtroom example, and think of the difference between 'we have found no conclusive evidence of the defendant's innocence', and 'we have found conclusive evidence of the defendant's guilt'. The latter is required in a well-run judicial system, for the reason Douglas cites: the very high cost of a mistaken guilty verdict.

Mapping this over to the cold fusion case – and remembering what was at stake – the question the DoE Panel should have answered is whether there was convincing evidence that there was no phenomenon of potential interest. Instead, it answered the easier one, concluding that there was no convincing evidence that there is a phenomenon. In a low-stakes case this would have been acceptable, of course – which suggests that the DoE Panel was blind to the distinction.

3.3 *Nature*

Melinda Baldwin's excellent (2015) history of *Nature* gives this account of the journal's role in the treatment of cold fusion in 1989–1990.

> Instead of being the forum where a new era of energy was declared, *Nature* quickly became a major center of cold fusion skepticism. By 29 March 1990, a year to the week after the first mention of cold fusion in *Nature,* [the Editor, John Maddox] felt

---

[5] In a piece in *Nature* reporting on the October EPRI-NSF meeting, David Lindley refers to the DoE panel as 'the DoE's "killer commission", as John Bockris of Texas A&M has dubbed it.' (Lindley 1989, 679) Bockris (1923–2013) was an electrochemist – according to McKubre (2017, 18) 'the father of modern physical electrochemistry and [an] experimental genius' – who had reported positive results in attempted replications of Fleischmann and Pons's work.

secure enough to declare "Farewell (Not Fond) to Cold Fusion" in the magazine's leader. (Baldwin 2015, 201)

As Baldwin goes on to say:

> During the cold fusion controversy, Maddox ... and the rest of the editorial staff cast the cold fusion episode as a battle between careful, peer-reviewed, properly conducted science and sloppy science revealed through press conferences in hopes of wealth through patents. Maddox wrote editorials criticizing Pons and Fleischmann's methods, associate editor David Lindley wrote news articles forecasting the death of cold fusion, and the journal's editorial staff gave significant space to cold fusion's most prominent scientific critics. Where *Nature* led, science reporters followed. News outlets such as *Time,* the *Economist,* and the *Wall Street Journal* all covered *Nature's* role in the cold fusion controversy and portrayed the journal's skepticism as proof that the scientific community was rejecting the Pons-Fleischmann claims. Ultimately, the cold fusion episode convinced many observers of the scientific journal's continued importance to the scientific community and illustrated *Nature's* influence among both scientists and laymen at the end of the twentieth century. (Baldwin 2015, 201–202)

We'll provide some examples relevant to the gatekeeping questions under discussion. In the issue marking the first anniversary of the 1989 press conference, David Lindley has an opinion piece called 'The embarrassment of cold fusion'. He closes with a remarkable recommendation about how science should police its borders.

> Perhaps science has become too polite. Lord Kelvin dismissed the whole of geology because his calculations proved that the Sun could be no more than a few million years old; Ernest Rutherford is still remembered for his declaration that talk of practical atomic energy was "moonshine" – but the stature of neither man has been noticeably diminished by their errors, which were as magnificent as their achievements. Kelvin and Rutherford had a common-sense confidence in the robustness of their judgements which the critics of cold fusion conspicuously lacked. Would a measure of unrestrained mockery, even a little unqualified vituperation, have speeded cold fusion's demise? (Lindley 1990, 376)

There are a number of puzzles here. Is this intended as a recommendation that both sides should be more vituperative? If so, then it is a recipe for a shouting match. Fleischmann was one of the most distinguished electrochemists of his day. Should he have resorted to mockery and vituperation, too?[6] He had a better claim to it, by Lindley's own standard's, than John Maddox or Lindley himself.

More likely, Lindley's advice is intended to be one-sided. It is only the respectable side who should be more vituperative. But then it has the 'buy low, sell high' problem. It can only be applied with the benefit of hindsight, when the issue is already resolved, one way or the other.

Most bizarrely of all, Lindley chooses examples in which the famous critics were simply *wrong.* If the great Rutherford could be mistaken about the practical relevance of atomic energy, this hardly inspires confidence in Maddox's own judgement, in the same issue:

> What has irretrievably foundered is the notion that cold fusion has great economic potential. The time has come to acknowledge that. It would be a cruel deception of a largely amused public not to admit that simple truth. (Maddox 1990, 365)

If Maddox had referred to cold fusion as moonshine, the irony here would hardly have been starker.[7]

3.4 *Nature*, DoE, and inductive risk

Ironies aside, we are interested in the question whether Maddox, Lindley and their *Nature* colleagues paid adequate attention to Douglas's issues of epistemic risk. Obviously, they had on the table the question whether cold fusion could be a practical source of energy, as Fleischmann and Pons believed and argued. They could hardly have failed to notice the relevance of this to the possibility of global warming caused by fossil fuels, in addition to the potential economic benefits of new sources of energy. From here, it doesn't take any great insight to see the potential costs of a false negative – of wrongly dismissing cold fusion. Did *Nature* consider this question? If so, we haven't found any sign of it. Maddox and his colleagues seem to have been blind to the issue.

To be clear, *Nature's* response is explicable, given the circumstances. Fleischmann and Pons announced their claims via press conference in a context of priority and patent anxiety, leaving many critical details incomplete or ambiguous. Subsequent technical clarifications were circulated informally, if at all, and often too late to guide early 'replication' attempts. In that

---

[6] Perhaps he did, at least in response to Lindley. Mallove (1991) reports as follows:

> Fleischmann and Pons shot back [at *Nature*] … In a blistering rebuttal they wrote: "In its extreme form, following the herd in editorial opinion is a manifestation of cultural fascism: the expression of convictions based on inadequately understood theories and facts. Scientific conformism is known as 'handle cranking' or 'me-too science.' Committee reports (which are editorials) are specialized ways of inducing scientific conformism. Electronic mail and fax machines are specialized ways of inducing scientific hysteria … If Lindley doesn't have the time to come now to Utah to gather information firsthand, then why doesn't he at least have the sense to use that well-known shortcut of establishing the scientific credentials of the believers and non-believers, namely the Citation Index."

By this point, relations between Fleischmann and Pons and *Nature* had been bad for many months, as we note below.

[7] In the interests of accuracy, we note that what Rutherford believed privately, in contrast to his public 'moonshine' comment, isn't entirely clear; see Jenkin (2011) for discussion.

setting, rapid public verdicts—both inside journals and in the wider press—became easier to form and harder to reverse.

Moreover, it appears that Fleischmann and Pons were not easy to deal with, from *Nature's* point of view, from an early point.[8] On May 1, 1989, just six weeks after the original press conference, *The Boston Herald* carried a news item from Associated Press, headed 'Researchers take issue with journal *Nature*' (Boston Herald, 1989).

> SALT LAKE CITY – Two scientists under attack for their claims of room-temperature nuclear fusion have lashed out at the source of the latest rebuff – the British scientific journal *Nature.* Fusion researchers B. Stanley Pons and Martin Fleischmann said the prestigious journal forced them to withdraw their paper and made inaccurate statements, and that one editor was "mumbling in the dark." …
>
> While laboratories around the world have duplicated the experiment and reported similar results, others have reported failures and critics have remained unconvinced. *Nature,* in an April 27 edition that contained several articles on fusion, speculated that the Utah experiment is fatally flawed and will never be verified.
>
> Pons said *Nature* had dipped to "sensationalism of the worst type. I have never in my life seen such a scandal by a scientific journal to discredit two people. They have a privileged position that they are using."
>
> David Lindley, *Nature*'s assistant physics editor, said there is frustration in the scientific world because the experiment hasn't been confirmed, and Pons and Fleischmann are inaccessible to researchers. (Boston Herald, 1989)

It is hard to find a justification for Pons' comment, at least in the April 27 issue of *Nature,* to which this piece refers. John Maddox's editorial appears under the heading 'What to say about cold fusion' (Maddox 1989). It is sceptical, but not entirely dismissive. He criticises Fleischmann and Pons on what seem reasonable grounds – failures to report sufficient details of their experiments, and especially the absence of control experiments – before concluding like this.

> So robust scepticism is the only wise view. There may be something in the Brigham Young phenomenon, but that requires careful confirmation. The Utah phenomenon is literally unsupported by the evidence, could be an artefact and, given its improbability, is most likely to be one.[9] (1989, 701)

---

[8] Beyond this, Fleischmann and Pons did blunder on their reported gamma spectrum associated with the excess heat phenomenon. Their attempt to remedy the error leaves much to be desired. See (Messinger 2023) for more discussion and reference to original correspondence in *Nature* between critics from scientists at MIT's Plasma Science and Fusion Center and Fleischmann and Pons.

[9] The phrase 'the Brigham Young phenomenon' refers to the claims of Steve Jones, whose paper is published in that issue of *Nature.* Philip Ball tells us that the *Nature* editorial office found Jones a great deal easier to deal with than Fleischmann and Pons. And while Jones's claims were comparatively more modest (e.g., neutron emission slightly above background but no excess heat), they were nevertheless extraordinary in conclusion: an apparently unknown solid-state enhancement of the fusion rate by tens of orders of magnitude. Why then the field so easily acquired pariah status, regardless of critiques of Fleischmann and Pons, remains an open question.

While Maddox does mention reputational factors in this piece, it is as a matter of the public view of science, not of Fleischmann and Pons specifically – he sees the entire episode as a threat to the reputation of science.

> No doubt the general opinion [about science] will depend on the outcome of attempts at replication, but the community might wish that its reputation did not hang on such a narrow thread, especially because the likelihood of replication fades as the days go by. (1989, 701)

Still, one can imagine a different role that *Nature* might have played, which – while emphasising the importance of 'careful, peer-reviewed, properly conducted science' (Baldwin 2015, 201) – also stressed the very high potential cost of a false negative, in this particular case. *Nature* editors could also have drawn on insights from the history and philosophy of science to note that many new scientific discoveries emerge from messy, contested episodes and prolonged ambiguity – which later textbook accounts commonly smooth into linear, hero-centered stories by retrospective selection and distortion (Kuhn 1970, Whitaker 1979).[10] All the more reason, then, to adopt a prudential approach, protecting the potential candle flame from the risk that it might be snuffed out prematurely.

*Nature* could have used its 'influence among … scientists and laymen' (Baldwin 2015, 202) to recommend such caution. While deploring science by press conference, it could have preached the risks of hasty dismissal, in a case in which so much was at stake. It would then have been appropriate to deplore, not promote, the attempt to resolve scientific disagreements by mockery and vituperation. But *Nature* did not show this caution. Instead, as Baldwin says, *Nature* became 'a major center of cold fusion skepticism' (2015, 201).

In particular, *Nature* bears some of the responsibility for the negative reputation accorded to cold fusion over the course of its first year – its pariah status, to use Goodstein's term. Why is this important? Because it has acted as a barrier to reopening the case. Already in 1994, as Goodstein says, Scaramuzzi's colleagues in Rome 'did not hear what he was saying', because he spoke from beyond the pale. As we'll see, this reputation has lasted a long time.

This illustrates a central issue for gatekeeping. In principle, scientific assessments are supposed to be provisional, always revisable if contrary evidence emerges. That's one of the great merits of the methodology. Yet it can only work in practice if challenges are actually possible – if challengers are able to make themselves heard. And nobody listens to pariahs.

These are difficult issues. To what extent, for example, should scientific gatekeepers ever avail themselves of epistemic slurs, such as 'crank' and 'pariah'? We will not discuss such questions here. Our point is just that cold fusion provides a striking historical example of the effects of reputational factors, which here intersect with issues of epistemic risk. It is one thing to fail to take proper account of the cost of a false negative, another to cement one's initial failure in place, by making it impossible to challenge.

---

[10] Douglas Allchin later makes a closely related point in the science-education literature, arguing that textbook 'discovery stories' often function as misleading pseudo-history and can foster distorted 'myth-conceptions' about how science advances (Allchin 2003, 2004). Similarly, interview studies by Wong and Hodson reveal practising scientists' accounts of research as heterogeneous, contingent, and socially embedded in marked contrast to the tidy, universalistic image of science typically projected in school curricula and textbooks (Wong & Hodson 2009, 2010).

In our view, Nature was guilty of both kinds of failure. It failed to pay proper attention to the risks of wrongly dismissing cold fusion, and it contributed to the reputational factors that made it so hard to appeal the case. For very good reasons, Nature is one of the most respected institutions in contemporary science. We are entitled to expect it to exercise its authority with care and responsibility. It failed to do so in this case, we suggest.

The first of these criticisms also applies to the DoE Panel, with equal or greater force.[11] Here we have the Report itself, and can be sure that the cost of a false negative is not noted explicitly. Yet this is a jury of experts, explicitly charged with a gatekeeping task. Perhaps the lesson to be drawn is that until an adequate understanding of epistemic risk is achieved among scientists themselves, such panels need expertise from outside science, to assist with their deliberations.

We emphasise that these criticisms do not depend on the benefit of hindsight. Behaviour can be reckless, even if no actual harm results from it. The early years of cold fusion represent a landmark failure of responsible gatekeeping, in our view, whatever the later history of the field.

## 4. 'Replication' and rush to judgment

Our criticisms above leant heavily on the issue of inductive risk. Would it be fair to say that if those factors could be set aside – if the planet already had abundant sources of clean energy, perhaps – the controversy in the early years was otherwise well-handled, as an episode in 'pure science'? Or are there gatekeeping lessons that do not depend on such factors? Our answers are 'no' and 'yes', respectively.

### 4.1 Historical context

Some prehistory. Fleischmann and Pons's work with deuterium-loaded palladium cathodes did not come out of thin air. Their experiments in the 1980s were embedded in a longstanding scientific context. It included both speculations about how solid state lattices might alter nuclear reactions, and empirical studies reporting anomalous effects in hydrogen-infused metals.[12]

As early as the 1920s, eminent researchers such as the chemist Friedrich Paneth, a former director of the Max Planck Institute for Chemistry, speculated that palladium may facilitate nuclear reactions of embedded hydrogen nuclei. In the mid-1970s, Friedwardt Winterberg, a theoretical physicist who had studied under Werner Heisenberg in Göttingen, proposed that piezoelectric crystals under stress could generate intense internal electric fields capable of accelerating deuterons to fusion-relevant energies. Winterberg later extended this idea to solid-state environments more generally, speculating that collective effects in a lattice – such as phonon coupling or electrostatic confinement – might enhance fusion rates in condensed matter systems.

---

[11] Unlike Nature, the DoE report does not mock cold fusion, so it is not guilty of the second failing in the same obvious way.
[12] For a useful survey of such work, see (Lewenstein & Baur, 1991).

Fleischmann was one of the world's leading electrochemists, with extensive expertise in the electrochemical loading of palladium and other materials. He also had a deep interest in systems far from equilibrium, and in the potential for extreme local conditions in electrochemical cells. And he was aware, of course, of the history of speculations and claims involving nuclear effects in metal hydrides. It was a natural progression for him to investigate such materials.[13]

4.2 The issue of 'replication'

Now to the question of replication. As we have seen, much of the early criticism of the work of Fleischmann and Pons turned on the claim that many attempts to *replicate* it had failed. What does 'replication' mean? In experimental science a genuine replication result, positive or negative, requires at least the following: access to a complete and reliable protocol for the original work; matching key experimental conditions (materials, methods, measurement systems); and deviation only where explicitly intended to test robustness.[14]

In the case of cold fusion, however, the vast majority of experiments conducted from March to October 1989 could not be replications in this sense. The experimenters concerned had no access to detailed protocols, and no direct guidance from the original team. Fleischmann and Pons published a short paper in the *Journal of Electroanalytical Chemistry* in April 1989 (Fleischmann Pons, 1989), but it focused primarily on calorimetric results and provided only partial descriptions of the experimental apparatus, materials, and procedures. Critical elements – including the precise method for palladium cathode preparation, deuterium loading dynamics, cell geometry, calibration routines, and control experiments – were either missing or ambiguously reported.

Additional technical information was not made widely available through standard venues, but was circulated informally, if at all, through private correspondence, press statements, and presentations – and generally arrived too late to inform early replication attempts. As Peter Hagelstein (MIT), one of the few researchers with early access to Fleischmann, noted in correspondence:

> It took on the general order of a year for groups to start weighing in with positive results, based on their versions of the experiment that were developed mostly in the absence of an appropriate technical description from Fleischmann and Pons. Probably the best early documentation is in their patent application which … became public about a year after the announcement, but even that is woefully short on the technical details one would like to have for a replication. (Hagelstein 2025)

---

[13] By 1989 Fleischmann had been working on palladium-deuterium systems for at least 40 years. As McKubre (2017, 18) says: '1948 was the year Martin first published on the Palladium–Deuterium system, and the year of my birth. This work has been going on for a very long time.' McKubre himself had worked with Fleischmann in Southampton in the 1970s, before the latter's cold fusion work. Peter Hagelstein (2025) reports that Fleischmann told him that in his experience as an electrochemist, he noticed that the calorimetry errors for palladium-deuterium systems were systematically higher than for palladium-hydrogen. This prompted him to try the cold fusion experiments.

[14] See Scott et al.'s (2022) 'To Err is Human; To Reproduce Takes Time', a recent editorial in the American Chemical Society journal *ACS Catalysis*, which makes a similar point. The authors stress that replicability presupposes detailed, accessible protocols and data (often lacking or incomplete), and emphasise that doing this kind of experimental work well is intrinsically time-consuming and often subject to human error.

Researchers eager to test the claims often resorted to improvised protocols and experimental configurations, leading to a wide array of divergent setups and outcomes. These efforts were widely labeled as 'replications', but they were not. They may have been rigorous in their own right, but they were methodologically disconnected from the original claim.[15]

Yet the rhetorical force of calling them 'failed replications' allowed the scientific establishment to dismiss the original claims, without confronting the ambiguity of the evidentiary base or the social dynamics of experimental reproduction. This dynamic reflects a broader epistemic problem: the premature framing of failure to replicate as failure of the phenomenon itself, without acknowledging the absence of a common experimental baseline.

The downstream effect of this misframing was profound: a rhetorical closure of inquiry that lasted decades, despite later experimental successes by groups with closer access to original protocols. It also shaped the institutional memory of cold fusion as a cautionary tale of error and embarrassment – when a more accurate diagnosis may be a breakdown in communication and epistemic coordination.

In fairness, this inattention to the importance of exact replication seems to have been a problem inside the LENR field, as well as outside it. In a 2008 piece called 'The importance of replication', Michael McKubre takes the field to task. He says that to his knowledge, there was only *one* exact replication of Fleischmann and Pons, in the important sense.

> Lonchampt *et al* [1996] set themselves the task of reproducing the original FPE work, in their words "*simply to reproduce the exact experiments of Fleischmann and Pons – to ascertain the various phenomena involved in order to <u>master the experiments</u>*". The phrase underlined is critical. Only from the point of mastery can systematic effects be studied, whether these are errors, artifacts or new physical processes.
>
> As far as I am aware Lonchampt and his team … remain the only group ever to attempt an exact engineering replication of the original Fleischmann Pons experiment. It helps considerably that they were engineers, not scientists. … They also were successful and closely specified the conditions under which replication was possible:
>
> > *i. The Fleischmann Pons calorimeter with precautionary measures taken is simple and precise.*
> > *ii. Their calorimeter is very accurate and well adapted to study cold fusion.*
> > *iii. The maximum error might be in the higher temperature range, and under any condition should not exceed 1% of the input power.*
> > *iv. The effect measured [now called the FPE] is:*
> > > *- below 70°C, between 0 and 5%*
> > > *- between 70°C and 99°C, about 10%*
> > > *- at boiling, up to 150%* (McKubre 2008, 13–14)

---

[15] (Hagelstein 2017) notes that some claimed attempts at replication were inadequate in a different way. He reports that in 1989 Fleischmann and Pons said that they found excess heat in 10–20% of their tested cathodes; whereas the celebrated MIT 'failed replication' tested just one cathode.

## 4.3 Rush to judgement

So the early history of cold fusion can be faulted, by contemporary gatekeeping standards, by the liberties it took with the notion of replication. These failings were compounded, of course, by the haste with which the scientific community at large rushed to judgement.

Was there a reason for such haste? A charitable view might return to the topic of inductive risk, and point out the risks of a false positive. Precisely because cold fusion claimed to be disruptive, with potential to revolutionize nuclear energy, a false positive would have been a major embarrassment to many parties. One way of understanding the response of *Nature,* the DOE panel, and the scientific ecosystem in general, is as simple caution. Imagine that cold fusion had attracted long term attention, with a sustained focus from the research community; and that the DOE and other science funders develop dedicated research programs to investigate the topic; and yet it eventually turned out to lead to nothing. This would have been a major blow to the institutions concerned, and to the credibility of science – a very public failure, for all concerned.

In the case of *Nature,* it might have been even worse. In her (2015) history of *Nature,* Melinda Baldwin notes there had been several prominent cases in which *Nature* had been taken to task for being too 'lenient' with controversial claims. In the 1970s, during the editorship of David Davies, *Nature* published a paper about extrasensory perception, and a sympathetic piece about the Loch Ness Monster. John Maddox returned to the editorship in the 1980s, and was of course aware of the controversy about those cases. Nevertheless, in June 1988, he published a piece by the French immunologist Jacques Benveniste, reporting results that, if confirmed, would have lent credibility to homeopathy. A sceptical editorial position did not save Maddox, or *Nature*, from considerable criticism.

So it is perhaps not surprising that a mere nine months later, when the opportunity presented itself, Maddox took up the chance to be hard-line with cold fusion. As we noted above (§3.4), he was certainly concerned that the episode would damage the public reputation of science. This is not to excuse what looks, in Douglas's words, like a failure to take into account the potential costs of a false negative – a failure greatly amplified, as we have said, by the derision that accompanied *Nature's* coverage. But it is to recognise some of the historical contingencies that may have played a role.

Such factors may have driven the felt need to say yea or nay to cold fusion, so hard and so fast – and to do so despite the ambiguities and uncertainties that we have seen to be acknowledged, at least to some degree, by Goodstein, the authors of the DoE report, and in the EPRI-NSF proceedings. But not everyone in the scientific community adopted this perspective. A striking contrary example is Julian Schwinger (1918–1994), the Nobel Laureate physicist of quantum electrodynamics fame.[16]

---

[16] Schwinger wasn't the only Nobel Laureate to show continuing interest in cold fusion. Willis Lamb, another famous quantum theorist, was another (Parmenter and Lamb, 1989). Together with Edward Teller, mentioned above (§2.2), these three seem the most distinguished proponents of what we here term the middle path.

## 4.4 The middle path (not taken)

At the time of the announcement of cold fusion, Schwinger was a professor at UCLA. In the last years of his career, and indeed until his death in 1994, Schwinger worked on formulating a theoretical mechanism for cold fusion effects. Schwinger is notable here not simply for the fact that someone of his calibre and theoretical background regarded cold fusion as sufficiently interesting to merit his time, but also for his ability to take seriously the *possibility* of cold fusion, without needing to come to a definitive verdict, one way or the other. This is how he put it in a brief paper released after his death.

> As Polonius might have said: "Neither a true-believer nor a disbeliever be." From the very beginning in a radio broadcast on the evening of March 23, 1989, I have asked myself—not whether Pons and Fleischmann are right—but whether a mechanism can be identified that will produce nuclear energy by manipulations at the atomic—the chemical—level. Of course, the acceptance of that interpretation of their data is needed as a working hypothesis, in order to have quantitative tests of proposed mechanisms. (Schwinger 1994)

Where others rushed to judgement, Schwinger saw the moment as an opportunity for some potentially fruitful work, and was able to maintain a wait-and-see attitude on the status of the Fleischman and Pons experiments. Regardless of the validity of the excess heat results, Schwinger saw an important research question to pursue: Is there a mechanism to 'produce nuclear energy by manipulations at the atomic—the chemical—level.' On the interest of these questions, Schwinger was full of conviction.

> My first attempt at publication, for the record, was a total disaster. "Cold Fusion: A Hypothesis" was written to suggest several critical experiments, which is the function of hypothesis. The masked reviewers, to a person, ignored that, and complained that I had not proved the underlying assumptions. Has the knowledge that physics is an experimental science been totally lost? The paper was submitted, in August 1989, to *Physical Review Letters.* I anticipated that *PRL* would have some difficulty with what had become a very controversial subject, but I felt an obligation to give them the first chance. What I had not expected—as I wrote in my subsequent letter of resignation from the American Physical Society—was contempt. (Schwinger 1994)

What influence, if any, did Schwinger's interest in the field have on the early history? He doesn't appear in the list of invitees to the EPRI-NSF meeting in October 1989. But he may have had some influence on another event that month. On 31st October, the DoE Panel held its last meeting, prior to the release of its final report. The meeting was eventful because the Co-Chair of the Panel, the physicist (and then a recent Nobel Laureate) Norman Ramsey, presented his colleagues with an ultimatum. He tendered a letter of resignation, but offered to withdraw it, if a preamble he had drafted could be added to the final report (with the intention, as he saw it, of softening the negative tone). This was agreed, with some minor modification to Ramsey's draft preamble.[17]

What has this to do with Schwinger? The other Co-Chair, John Huizenga, later described these events like this:

> It would have been damaging to the impact of our final report had Ramsey, a new Nobel Prize winner in physics, resigned as our final report was just being completed. Facing this hard choice, three hours before our panel was to disperse, the panel opted to accept Ramsey's preamble, rather than his resignation … (Huizenga 1992, 91)

Huizenga proposes an explanation of Ramsey's actions, and this is where Schwinger plays a role.

> In retrospect, Ramsey's insistence on including the preamble may have been in deference to two colleagues, both Nobel Prize winning theorists in physics. Ramsey had received communications, already prior to completing our interim report, from Professors Julian Schwinger (UCLA) and Willis E. Lamb. Jr. (University of Arizona) who had invented 'explanations' of cold nuclear fusion in an atomic lattice. (Huizenga 1992, 92)

Huizenga goes on to note with satisfaction that '[w]hile the panel was willing to accept the above caveat [i.e., Ramsey's preamble], which weakened the report somewhat, it insisted on preserving the report in its entirety, including a forceful concluding item as given above by conclusion five' (1992, 92). As we noted above, conclusion five says this:

> Nuclear fusion at room temperature, of the type discussed in this report, would be contrary to all understanding gained of nuclear reactions in the last half century; it would require the invention of an entirely new nuclear process.

Huizenga seems to have regarded this as the killer conclusion of his 'killer commission' (cf. n. 5 above). Yet 'an entirely new nuclear process' is an apt description of what Schwinger and Lamb

---

[17] This is the Preamble as published in the final report:

> Ordinarily, new scientific discoveries are claimed to be consistent and reproducible; as a result, if the experiments are not complicated, the discovery can usually be confirmed or disproved in a few months. The claims of cold fusion, however, are unusual in that even the strongest proponents of cold fusion assert that the experiments, for unknown reasons, are not consistent and reproducible at the present time.
>      However, even a single short but valid cold fusion period would be revolutionary. As a result, it is difficult convincingly to resolve all cold fusion claims since, for example, any good experiment that fails to find cold fusion can be discounted as merely not working for unknown reasons. Likewise the failure of a theory to account for cold fusion can be discounted on the grounds that the correct explanation and theory has not been provided. Consequently, with the many contradictory existing claims it is not possible at this time to state categorically that all the claims for cold fusion have been convincingly either proved or disproved. Nonetheless, on balance, the Panel has reached the following conclusions and recommendations.

had proposed – that is, an entirely new application to nuclear physics of the existing quantum theory, in which they were among the leading theorists of their day. Between the interim report in July and the final version in early November, Schwinger had had his public falling-out with the American Physical Society. Huizenga surely knew of those events. It must have taken considerable hubris to regard conclusion five as a Road Closed sign, rather than an invitation to further work – work of precisely the kind that Schwinger and Lamb had in mind.

Mehra and Milton's (2000) biography of Schwinger has a detailed account of his interest in cold fusion. In the present context, it is interesting, among other things, for its framing of the field. After describing the famous news conference, Mehra and Milton continue like this.

> [A]lthough the subject continued to receive serious attention for the rest of 1989, by the summer of that year most believed it was pathological at best, fraudulent at worst. This became confirmed a year later, when the cold fusion conferences were effectively closed to all but true believers …
>
> So it was a shock to most physicists when Schwinger began speaking and writing about cold fusion, suggesting that the experiments of Pons and Fleischman were valid, and that the palladium lattice played a crucial role. In one of his later lectures on the subject in Salt Lake City, Schwinger recalled, 'Apart from a brief period of apostasy, when I echoed the conventional wisdom that atomic and nuclear energy scales are much too disparate, I have retained my belief in the importance of the lattice.' (Mehra and Milton 2000, 549–550)

Veterans of that time might be wryly amused to learn that they were considered responsible for their own isolation, in allegedly closing their meetings to outsiders. Did they label themselves pariahs, too? In any case, the direction of causation aside, it is simply not true. Some of the leading mainstream observers of cold fusion, including Huizenga himself, continued to attend ICCF meetings for some years. (He was certainly present at the 1990 meeting in Salt Lake City, at which Schwinger gave the lecture from which Mehra and Milton are quoting here.)

Later in the chapter, Mehra and Milton sum up like this:

> By [the early 1990s], even [Schwinger] must have seen that the evidence for cold fusion could no longer be taken seriously. We must ask why he was so eager to jump on the cold fusion bandwagon, when most physicists approached the matter cautiously. The answer, in large part, may be found in his own experience. The rather 'contemptuous dismissal' of his source-theory program by most of the field theory community made him sensitive to the plight of the underdog. He always insisted on understanding things his own way. His way was often called conservative, yet this … is a great oversimplification. His openness to the cold fusion hypothesis may have been unfortunate, but was completely consistent with his attitude to science. As he concluded his Sigma Xi lecture on 'Conflicts in physics' with a quotation from Boltzmann: 'Who has seen the future? Let us have free scope for all directions in research; away with all dogmatism.' … Schwinger continued to hold his open-minded view of cold fusion to the end. (Mehra and Milton, 553)

This is interesting for its perspective on Schwinger – confirming the open-mindedness we praised above – and again for its view of the field, from Mehra and Milton's perspective. Was cold fusion ever a bandwagon? If so, then only briefly, surely, before it became a *tombereau,* conveying its cargo into exile, through a noisy and contemptuous crowd. This version of the metaphor fits much better with Mehra and Milton's stress on Schwinger's open-mindedness, and suspicion of popular dogmatism. Full credit to him that he was willing to climb on board.

Mehra and Milton's biography of Schwinger was published in 2000, a decade after the events of 1989. There is no sign that they were aware of work in the field in that period that might have led them to a more nuanced view of Schwinger's interest in it. As we'll see in a moment, such work was certainly there, for those with a motivation for taking a look. But it was hidden from sight by the reputation trap. We don't know whether Mehra and Milton ever considered the possibility that Schwinger might have been right.[18] But it would have been risky to say so, if so. They would have exposed themselves to the same ridicule as Schwinger himself, without the protection of a Nobel Prize.

## 5. The Middle Ages (1995–2014)

We are going to pass very briefly over the two decades 1995–2014. We will comment on two things: (1) a second DoE report in 2004; and (2) two examples of what the field looked like at this point, both in the public perception and to anyone taking a serious look (these being very different, as we'll see). One of our two examples is a development in Japan in 2012, which will play a significant role in the next part of the story.

### 5.1 The second DoE report

At the request of researchers within the field, the DoE conducted a second review of LENR in 2004. A survey paper prepared by Peter Hagelstein, Michael McKubre and others (Hagelstein et al 2006) was reviewed by eighteen panellists, from several disciplines. Nine of the reviewers also participated in a one-day meeting at McKubre's lab at SRI in Menlo Park, California.

The LENR community was disappointed with this review, but it was certainly not wholly negative. Reviewing its conclusions in 2021, DoE Fellow Dr Katharine Greco notes that the 2004 DoE panel was 'nearly unanimous' that

> funding agencies should entertain individual, well designed-proposals [into]
> – Whether … there is anomalous energy production in Pd/D systems
> – Whether … D-D fusion reactions occur at energies ~eV (Greco 2021)

Some LENR researchers complain that this recommendation was not actually followed – that funding proved as elusive as before (Maguire 2014). For our purposes, however, what's relevant is that DoE's panel of eighteen skeptical reviewers did not simply dismiss the field. Their individual views varied. McKubre reported that the nine who participated in an in-person meeting at his lab were almost uniformly positive (Maguire 2014). The report itself states that on the question of 'experimental evidence for the occurrences of nuclear reactions in condensed matter at low energies (less than a few electron volts)', one third of the reviewers were either completely convinced or 'somewhat convinced' (DoE 2004).

---

[18] There is no sign of it in (Milton 2019), a contribution to a centenary conference for Schwinger the preceding year.

Clearly, this is far from the received view, at that time and later, that cold fusion was disreputable pseudoscience – something that should be shunned by respectable researchers. The 2004 DoE report does not appear to have had any influence on this reputation, but it does nothing to support or justify it. But nor, on the other hand, does it do anything to correct the astonishing disregard of the prudential factors. Like its predecessor in 1989, the 2004 panel seems to have been blind to such considerations.

5.2 Views from outsiders (I): *60 Minutes*

We want to convey an impression both of general attitudes to the field in this period, and of what it looked like to some of those who, for one reason or another, took a closer look. In both respects, one useful source is an episode of the CBS program *60 Minutes* titled 'Cold Fusion is Hot Again', aired on April 19, 2009. We quote here from the transcript of that piece that is still to be found on the CBS website (CBS 2009).

> *60 Minutes* turned to an independent scientist, Rob Duncan, vice chancellor of research at the University of Missouri and an expert in measuring energy.
> "When we first called you and said 'We'd like you to look into cold fusion for *60 Minutes*,' what did you think when you hung up the phone?" [Scott] Pelley asked Duncan. "I think my first reaction was something like, 'Well, hasn't that been debunked?'" he replied.

The transcript describes how *60 Minutes* asked Duncan to go to Israel, to visit a lab that claimed successful experiments producing excess heat. Duncan spent two days there, examining the experiments and the accuracy of the measurements.

> Asked what he thought when he left the Israeli lab, Duncan told Pelley, "I thought, 'Wow. They've done something very interesting here.'" He crunched the numbers himself and searched for an explanation other than a nuclear effect. "I found that the work done was carefully done, and that the excess heat, as I see it now, is quite real," Duncan said. Asked if [he] was surprised that he'd hear himself saying that, Duncan told Pelley, "Very much. I never thought I'd say that."

*60 Minutes* then asked Duncan about the reputation of the field.

> "You know, I wonder how you feel about going public endorsing this phenomenon on *60 Minutes* when maybe 90 percent, I'm guessing, of your colleagues think that it's crackpot science?" Pelley asked.
> "I certainly was among those 90 percent before I looked at the data. And I can see where they'll be very concerned when they see this piece. All I have to say is: read the published results. Talk to the scientists. Never let anyone do your thinking for you," he replied.

Duncan also commented on the potential of the process to be a useful source of energy.

> If you ask me, is this going to have any impact on our energy policy, it's impossible to say, because we don't fundamentally understand the process yet. But to say, because we don't fundamentally understand the process and that's why we're not going to study it, is like saying, 'I'm too sick to go to the doctor.'

A few days after the *60 Minutes* episode, Duncan was an invited speaker to the Missouri Energy Summit at the University of Missouri, where he was Vice Chancellor for Research. He described his experiences and conclusions in making the *60 Minutes* program, and commented on the reputational factors associated with the field.

> This negative reaction resulted in what I consider to be a loss … of objectivity. When I went on the *60 Minutes* piece, I was contacted by a highly prominent professor from an Ivy League university, who was just really …  angry with me. For having done the piece. And the point was, I laid out the scientific case [to the prof on the phone] but he flatly wouldn't consider it. And when I said, 'come on, why don't you just work with me here, through the data,' he said essentially, 'well, you know, us high caliber physicists have done that before, and there has never been anything there. So you charlatans just can go on and do whatever you like.' Okay. Well. It is interesting: my scientific reputation – I guess, at least to him – had been stronger before I did the piece.
> But now, the point is, real science, possibly with outstanding engineering consequences, suddenly becomes a pariah science. A science where no one can go … (LENR-CANR.ORG 2009, Duncan 2009)

Duncan put his career where his mouth was, and went on to work on cold fusion himself, initially at Missouri (see §6.1 below) and later at Texas Tech University.

5.3 Views from outsiders (II): Clean Planet

Hideki Yoshino is a Japanese businessman. In 2012 he founded Clean Planet Inc., now one of the interesting companies in the commercial cold fusion world. We will mention the recent activities of Clean Planet in the next section. For now, we use it as a second indication of what the field looked like in the period 1995–2014, to those with a motivation to take a look. The following account is from a Forbes Japan article in August 2023.

> Hideki Yoshino, CEO of Clean Planet, founded the English conversation school GABA while studying … at the University of Tokyo. After getting the business on track … he obtained a master's degree from London Business School, and then worked as an environmental investor around the world.
> At that time, the Great East Japan Earthquake occurred. Yoshino, seeing the runaway nuclear reaction at the Fukushima Daiichi Nuclear Power Plant, realized that "a completely new clean energy is needed for a sustainable future." He returned to Japan from overseas immediately and used his own network to gather information from relevant ministries and agencies to international societies of advanced physics.

When he attended the ICCF (International Conference on Cold Fusion) held in Korea [in 2012], he was attracted to "cold fusion" as the next generation of energy. He wanted to use this power to bring an energy revolution from Japan … to the world. With that in mind, he founded Clean Planet in September 2012.

Yoshino's turning point was when he met Dr. Yasuhiro Iwamura, who was then researching condensed matter nuclear reactions at Mitsubishi Heavy Industries. The two men agreed on the concept of "New Energy, New Future (Let's open up the future of humanity by inventing new energy sources)."

In 2015, Clean Planet launched the "Condensed Matter Nuclear Reaction Joint Research Division" with Tohoku University … and Iwamura was appointed as a specially appointed professor at Tohoku University. Under the industry-academia collaboration system, the first step towards full-scale research and practical application of next-generation energy has begun. (Hatakeyama 2023)

As we say, this history is evidence of the credibility of the field within Japan in 2012, at least to someone with a motivation to take a close look at it.[19] And Clean Planet appears to have been very successful – more on this below.

## 6. The Renaissance (2015–2024)

Now to the recent history. The last ten years have seen some remarkable changes in the field, both reputational and (apparently) technological. In the former case, as we'll see, much of the credit belongs to a very familiar name.

### 6.1 The private landscape

In September 2016, *New Scientist* published an article by the physicist and science writer Michael Brooks entitled 'Cold fusion: science's most controversial technology is back', noting increasing interest in the field, especially from private investors.

> Fast forward 25 years [from the rejection of the 1989 claims], and thaw is in the air. You won't hear the words "cold fusion", but substantial sums of money are quietly pouring into a field now known as low-energy nuclear reactions, or LENRs. (Brooks 2016)

Brooks quotes a number of researchers in the field, including Dr Graham Hubler, the director of the Sidney Kimmel Institute for Nuclear Renaissance at the University of Missouri in Columbia (SKINR), 'a cold-fusion lab established in 2012 with $5.5 million of philanthropic funding.'

---

[19] By recruiting and partnering with established academic LENR researchers, Clean Planet drew on an academic lineage in Japan of materials-driven, incrementally refined experimentation across a small number of enduring experimental threads in metal–hydrogen/deuterium systems. See Kasagi & Iwamura (2008) for a contemporaneous overview; and Arata & Zhang (1999, 2002), Kasagi et al. (2002), Kitamura et al. (2018), and Iwamura et al. (2024) for representative peer-reviewed examples.

(Brooks 2016) SKINR was founded by Rob Duncan, whose 'conversion' to the cause of cold fusion by *60 Minutes* we described above. Hubler himself had had a forty year career in naval research labs, with a longstanding active interest in cold fusion. (Basi, 2013; Hubler 2015) Brooks quotes Hubler as saying, 'We're convinced there's some sort of energy source here, I wouldn't have taken this job if I didn't feel that way.'

As if to hedge its bets in tackling such a controversial topic, the same issue of *New Scientist* includes a dismissive editorial, reproducing some of the old reputational tropes.

> Not many respectable scientists would touch cold fusion today. But a few do, and so it has never really gone away. Private funding, mostly from investors, has now reached millions of dollars. Alerted, the US Congress has asked for a report on the case for public funding. … [I]f there is anything to cold fusion, it would be in the public interest for it to be investigated properly. But that's an enormous if. There's still no compelling reason to think cold fusion will work. Let those with money to burn take the risk and, if proven right, the rewards for their chutzpah too. For the rest of us, cold fusion is better off left out in the cold. (Anonymous 2016)

This repeats some of the old mistakes of *Nature* and the DoE report, in our view. Given the likely costs of a false negative, the relevant question isn't whether there are compelling reasons to think that cold fusion will work, but whether there are compelling reasons to think that it will *not* work. It is foolish to create artificial, sociological risks for those who seek to answer the latter question, by treating them as pariahs.[20]

Sneering aside, the suggestion that the onus for proving the broader scientific community wrong should lie on private investors, presumably working in secret in the hopes of ultimately capitalizing on their findings, is also unhelpful. This preference for working in secret, together with the notion that a breakthrough is just within reach by a small organization alone, is shared, unfortunately, by many of the staunchest cold fusion advocates.[21] But this is rarely how good science is done, normal or not. Science is a sociological phenomenon, requiring communities of researchers, meetings where ideas are shared and constructive scrutiny is provided, and even healthy competition. If one accepts (as *New Scientist* is doing here, at least nominally) that the dismissal of cold fusion may have been a false negative, it seems foolish to think that the remedy lies in a private, secrecy-first approach.

Returning to Brooks's article, it also mentions work in Japan:

> Post-Fukushima, Japan has also seen a wave of interest in LENR for energy generation, with Mitsubishi, Toyota and Nissan all investing money. Last year, the Japanese government's New Energy and Industrial Technology Development Organization announced a programme called "Energy and the Environment New Leading

---

[20] As (Price 2019) puts it, '[w]e desperately need some new alternatives to fossil fuels.' So we need clever people to look for them, 'even in unlikely corners.' As he says, that's simple prudence. In other words, he concludes, 'we need some unconventional thinkers … and we need to cheer not sneer at their efforts.'

[21] As McKubre (2019, 3) puts it: 'In my experience experimenters and their backers in the CMNS field with strong positive results are very reticent to publish or even discuss or disclose their results in public.'

Technology" that called, among other things, for research into technologies that induce heat reactions between metals and hydrogen.

Brooks also mentions Clean Planet, whom we encountered above:

> Hideki Yoshino, a language schools magnate, has set up a company called Clean Planet to research "cleaner, safer, and more abundant resources such as solar, geothermal, LENR (also known as cold fusion), and wind to supply our energy needs."

.

Nearly a decade later, Clean Planet appears to be doing very well. While it is difficult to get an overview of the current commercial landscape,[22] Clean Planet appears to be among the leaders. In 2021 it announced a collaboration with Miura, one of Japan's largest manufacturers of industrial boilers, to develop prototype LENR-based boilers for industrial purposes (Miura 2021).[23] Clean Planet also has equity funding both from Miura and from the Mitsubishi Corporation (Clean Planet 2022). In 2025, it was awarded 1 billion yen (US$6.8m) from the Tokyo Metropolitan Government, from funding the city's Zero Emission policies (Wade 2025).

Clean Planet has attracted some attention in the Japanese business press (e.g., Hatakeyama, 2023), but little in the West. The contrast with coverage of small reported advances towards thermonuclear fusion (i.e., 'hot' fusion) is striking. It is an indication that the field is still treated with suspicion, in some quarters.

## 6.2 Google enters the chat

Late in 2014, two senior engineers at Google, Ross Koningstein and David Fork, published a paper analyzing our civilization's energy future, in the light of the constraints of global warming (Koningstein and Fork 2014). They concluded that known renewable energy sources were going to be insufficient, and hence that new clean energy sources must be found. Michael McKubre takes up the story:

> This article appeared in *IEEE Spectrum* in November 2014. Importantly, and before that, rather than congratulating themselves on their analysis and conclusions, the authors set out with Google's support to address that perceived need. … Google saw a problem, saw a potential solution, enlisted support and set out to do something about it. (McKubre 2019, 1)

One part of the potential solution was to re-evaluate cold fusion. As Koningstein (2024) describes it, the proposal emerged in discussion between himself and the venture capitalist and incoming Google Quantum COO, Matt Trevithick.[24]

---

[22] Reputational factors may still provide a motive for confidentiality, in addition to ordinary commercial reasons.
[23] Miura's own announcement (Miura 2021) mentions Google's interest in the field (see below) – reputational cover at work again.
[24] Trevithick himself describes the Google program in (Trevithick 2021).

> Trevithick had been scouting for scientists who were open to the idea that unusual states of solid matter could lead to cold fusion. Google greenlit the program and recruited Trevithick to lead it, and we ended up funding about 12 projects that involved some 30 researchers. (Koningstein 2024)

The program was kept confidential until 2019, when, to the surprise of many, *Nature* published a Perspectives piece by some of these Google-funded researchers (Berlinguette et al 2019). This is how they present their work:

> The 1989 claim of 'cold fusion' was publicly heralded as the future of clean energy generation. However, subsequent failures to reproduce the effect heightened scepticism of this claim in the academic community, and effectively led to the disqualification of the subject from further study. Motivated by the possibility that such judgement might have been premature, we embarked on a multi-institution programme to re-evaluate cold fusion to a high standard of scientific rigour. Here we describe our efforts, which have yet to yield any evidence of such an effect. Nonetheless, a by-product of our investigations has been to provide new insights into highly hydrided metals and low-energy nuclear reactions, and we contend that there remains much interesting science to be done in this underexplored parameter space.
>
> So far, we have found no evidence of anomalous effects claimed by proponents of cold fusion that cannot otherwise be explained prosaically. However, our work illuminates the difficulties of producing the conditions under which cold fusion is hypothesized to exist. This result leaves open the possibility that the debunking of cold fusion in 1989 was perhaps premature because the relevant physical and material conditions had not (and indeed have not yet) been credibly realized and thoroughly investigated. Should the phenomenon happen to be real (itself an open question), there may be good technical reasons why proponents of cold fusion have struggled to detect anomalous effects reliably and reproducibly. Continued scepticism of cold fusion is justified, but we contend that additional investigation of the relevant conditions is required before the phenomenon can be ruled out entirely. (Berlinguette et al 2019)

Later in the piece, they conclude like this:

> *Call to action*. Fusion stands out as a mechanism with enormous potential to affect how we generate energy. This opportunity has already mobilized a 25 billion dollar international investment to construct ITER. Simultaneous research into alternative forms of fusion, including cold fusion, might present solutions that require shorter timelines or less extensive infrastructure.
>
> A reasonable criticism of our effort may be 'Why pursue cold fusion when it has not been proven to exist?'. One response is that evaluating cold fusion led our programme to study materials and phenomena that we otherwise might not have considered. We set out looking for cold fusion, and instead benefited contemporary research topics in unexpected ways.

> A more direct response to this question, and the underlying motivation of our effort, is that our society is in urgent need of a clean energy breakthrough. Finding breakthroughs requires risk taking, and we contend that revisiting cold fusion is a risk worth taking. (Berlinguette et al 2019)

Three comments. First, *not* revisiting cold fusion is risk-taking, too, and potentially a more serious one. That's the feature that this case shares with the more obvious cases of low probability high-impact risks – the high potential cost of a false negative. Second, for the risks that (Berlinguette et al 2019) have in mind, the degree of risk depends on sociological factors. Do researchers put their own careers and reputations at risk? If so, we can do something about it, by pushing back against the reputation trap. The organisers of the Google program were well aware of this, of course, and seem to have seen the program, amongst other things, as an exercise in reputational engineering – an outstandingly successful one, as we'll see.

Our third comment is from a different angle. The Google-funded teams arguably fell into the same epistemic trap that derailed the field's early history: they framed their experimental campaigns as *replications* of earlier cold fusion claims, even though the original experimental protocols they sought to reproduce were never specified in sufficient detail to permit strict replication. As in 1989, many of the key variables – including material microstructure, electrode treatment, loading dynamics, and system history – remain underspecified or unavailable, often known only tacitly to the original experimenters.

Without access to such information, the Google team's efforts may be more accurately characterized as *independent approximations* or *targeted reconstructions*, rather than genuine replications in the classical scientific sense. Yet by describing them as replications, the researchers risked reinforcing a rhetorical frame that equates *failure to observe* with *falsification* (when what was actually demonstrated may also have been a mismatch in implementation). This ambiguity continues to obscure the epistemic landscape of the field. It suggests that the lessons of 1989 – particularly the importance of distinguishing between negative results and failed reproductions under uncertain protocols – remain to be fully digested, even by those most committed to restoring the field's scientific credibility.

6.3 Google's reputational influence

What impact did Google's support have on the reputation of LENR? Some felt that *Nature's* own initial reaction was unpromising. The same issue of *Nature* contained an editorial and a commentary piece concerning the Google work. The latter was by the science writer Philip Ball, himself a former *Nature* editor, and was bylined like this:

> Why revisit long-discredited claims for a source of abundant energy, asks Philip Ball? Because we are still learning how to treat pathological science. (Ball 2019a, 601)

That tells *Nature's* readers that cold fusion is pathological science, and long-discredited. As in the case of the *New Scientist* editorial we mentioned above, this byline has the tone of someone who hasn't yet come to terms with the fact that their earlier hostile judgement may have been too hasty.

However, things were improving. A few months after the initial announcement, Ball spoke to some of the scientists involved in the Google-funded work. Here he quotes Curtis Berlinguette, head of the Berlinguette research group at UBC, Vancouver, and lead author on the Perspectives piece.

> "Renewable energy and fusion technologies are not scaling at the pace we need them to," says Berlinguette. "If cold fusion were realizable, it could take the world into an era of energy surplus rather than scarcity. It therefore seemed irresponsible to not take another look at it. For me, cold fusion started in 2015," says Berlinguette. "Prior to that, I didn't know enough about it to have an opinion. I was driven simply by curiosity to learn more about the field." (Ball 2019b, 883)

There is much to like here, from our point of view – especially Berlinguette's ability to look beyond the reputational factors to the true imperatives of the case ('It … seemed irresponsible not to take another look at it'). His open-minded epistemic attitude is also welcome.

Berlinguette's webpage soon said things like this: 'His program also likes to work on high risk, high impact clean energy projects like cold fusion' (UBC 2021). This provides welcome cover for younger scientists to work on these topics, without such a risk to their own careers.

Did Google's reputation provide cover for other research groups in the field? The answer is a resounding 'yes'. We'll offer two examples. The first is from 2020. In that year, the EU's Horizon 2020 funding program awarded two large grants to multi-institution teams working on LENR. One, the HERMES project, was funded at almost €4m over the five years 2020–2024. In a description of the project on the funder's website, the leverage of the Google work is explicit.

> [C]old fusion remains a controversial topic in the scientific community. … Since 2015, Google has been funding experiments into cold fusion. …The EU-funded HERMES project will employ advanced techniques and tools developed over the last few decades to investigate anomalous effects of deuterium-loaded palladium at room and intermediate temperatures. (CORDIS 2021a)

The second EU grant was for the CleanHME project, funded at €5.7m over 2020–2024. It was coordinated from Szczecin, Poland, and was described like this:

> With climate change being a major global concern in recent times, new efficient clean energy sources are in high demand, and there has been a rise in the use of many of them, such as solar or wind generators. One very promising energy source is hydrogen–metal energy (HME), which could be used for small mobile systems as well as in stand-alone heat and electricity generators. Unfortunately, little research has been conducted concerning HME. The EU-funded CleanHME project aims to change this. It will produce an elaborate comprehensive theory of HME phenomena that would assist in the optimisation of the process and construct a compact reactor to test HME technology. (CORDIS 2021b)

CleanHME appears to have been by far the more productive of the two projects.[25] Among other things, their website notes collaboration with new US-based projects, to which we now turn.

6.4 ARPA-E

We have already described the US DoE reports of 1989 and 2004. The Advanced Research Projects Agency-Energy (ARPA-E) is DoE's version of the well-known Defense Advanced Research Projects Agency (DARPA). ARPA-E says that it 'advances high-potential, high-impact energy technologies that are too early for private-sector investment'. (ARPA-E 2021a)

In October 2021 – under the influence, apparently, of Google's work – ARPA-E hosted a workshop on LENR. It was described as follows:

> The objective of this workshop was to explore compelling R&D opportunities in Low-Energy Nuclear Reactions (LENR), in support of developing metrics for a potential ARPA-E R&D program in LENR. Despite a large body of empirical evidence for LENR that has been reported internationally over the past 30+ years in both published and unpublished materials, as well as multiple books, there still does not exist a widely accepted, on-demand, repeatable LENR experiment nor a sound theoretical basis. This has led to a stalemate where adequate funding is not accessible to establish irrefutable evidence and understanding of LENR, and lack of the latter precludes the field from accessing adequate funding. Building on and leveraging the most promising recent developments in LENR research, ARPA-E envisions a potential two-phase approach toward breaking this stalemate: (1) Support targeted R&D toward establishing at least one on-demand, repeatable LENR experiment with diagnostic evidence that is convincing to the wider scientific community (focus of this workshop); (2) If phase 1 above is successful (metrics to be determined), support a broader range of R&D activities (to be defined later) toward better understanding of LENR and its potential for scale-up toward disruptive energy applications, thus setting up LENR for broader and more systematic support by both the public and private sectors. (ARPA-E 2021b)

This workshop was more than a symbolic gesture; it laid the groundwork for the most substantial U.S. government investment in the field since the early 1990s.[26] In 2022, the agency followed through with a formal Funding Opportunity Announcement (FOA DE-FOA-0002784) and launched an Exploratory Topic in LENR research, providing approximately $10 million in funding across eight multidisciplinary projects at universities, a national laboratory, and small businesses.

This initiative emerged from ARPA-E's sober recognition that the field remained caught in a self-reinforcing epistemic impasse. As ARPA-E's Teaming Partner Announcement put it:

---

[25] Among their results, a recent paper in a leading physics journal reports experimental signatures indicative of a new channel of the deuteron-deuteron reaction at very low energy (Dubey, R. et al. 2025).
[26] See Metzler (2021) for a perspective on the field presented at this workshop.

> LENR as a field remains in a stalemate where lack of adequate funding inhibits the rigorous results that would engender additional funding and more rigorous studies. (ARPA-E 2022)

The aim of the new program was to cut through this loop by establishing, once and for all, whether on-demand, repeatable LENR experiments with nuclear diagnostics could be achieved – or, conversely, whether the phenomenon could be ruled out decisively.

The Teaming Partner Announcement laid out the stakes with unusual clarity:

> Based on its claimed characteristics, LENR may be an ideal form of nuclear energy with potentially low capital cost, high specific power and energy, and little-to-no radioactive byproducts. If LENR can be irrefutably demonstrated and scaled, it could potentially become a disruptive technology with myriad energy, defense, transportation, and space applications. (ARPA-E 2022)

To move beyond the replication-centric framing that has hampered past research, ARPA-E emphasized hypothesis-driven design over open-ended reproduction. In its words:

> This forthcoming ARPA-E Exploratory Topic program aims to build on recent progress in the field, with strong emphases on testing/confirming specific hypotheses … and an insistence on peer review and publication in top-tier journals. (ARPA-E 2022)

The selected projects span a variety of technical approaches – from nanoparticle-mediated experiments and gas cycling systems to advanced nuclear diagnostics – but they share a commitment to systematic, high-rigor investigation. Importantly, the initiative also funded capability teams, on nuclear radiation diagnostics at the University of Michigan and on materials analysis at Robert Duncan's lab at Texas Tech – ensuring that diagnostics are held to the same standards as those used in more established branches of science.

The overarching message was clear: LENR might be controversial, but it is no longer off-limits. As ARPA-E Director Evelyn Wang put it in the program's official press release: 'While others have shied away from this space, ARPA-E wants to break through the knowledge impasse and deepen our understanding.' (ARPA-E 2023)

## 6.5 The quest for a reference experiment

One of the most consequential strategic decisions in the design of the ARPA-E LENR Exploratory Topic was its emphasis on real-time nuclear diagnostics – such as neutron, gamma, and charged particle detection – over the more historically prominent focus on excess heat measurements. This pivot was not accidental. It was the outcome of sustained deliberation following the 2021 workshop and internal agency discussions aimed at identifying a pathway that could credibly address the central epistemic bottleneck: Is there a real, replicable physical phenomenon here, or not?

In the decades since 1989, hundreds of experimental LENR reports have claimed the observation of anomalous heat – calorimetric results that suggest more energy output than can be accounted for by chemical processes alone. But calorimetry, while potentially powerful, is also subtle, indirect, and methodologically fragile. It is vulnerable to accusations of measurement error, baseline drift, or misinterpretation of system losses. And because it does not directly indicate a nuclear process, excess heat – however well measured – often fails to satisfy the skepticism of nuclear physicists trained to demand nuclear signatures as the *sine qua non* of nuclear reactions.

In contrast, real-time radiation measurements – especially if backed by independent diagnostics and well-characterized backgrounds – offer an avenue for more irrefutable, physics-level evidence. Though such results have historically been rarer in the LENR literature than calorimetric ones, there are still dozens of reports that document anomalous radiation: low-level neutron bursts, unexpected gamma emissions, and charged particle tracks.[27] However, these findings have tended to be episodic, under-instrumented, or non-reproducible, leaving them open to dismissal. The ARPA-E program was designed to change that. Early reports (Karahadian et al 2025) suggest that the program succeeded in that ambition.

In its planning documents and communications, the agency made clear that what the field lacked was not so much a quantity of suggestive results, but a single, unambiguous, community-credible 'reference experiment' – a protocol that could be reproduced across labs and demonstrate a nuclear-level anomaly with clarity. To this end, ARPA-E encouraged applicants to prioritize radiation diagnostics in their experimental designs. While calorimetry was still permitted – particularly where it complemented nuclear measurements – it was demoted in strategic priority.

These decisions represent a clear break from past practice in the field. It shows that ARPA-E is not simply revisiting the claims of 1989 but recasting the evidentiary priorities to fit the needs of a rigorous, early-stage program tasked with settling the 'Is it real?' question. In doing so, it may also shift the sociology of LENR research itself, redefining what counts as a valid signal – and ultimately, what might count as scientific vindication.[28]

## 7. Two forks

We now want to make some comments, partly speculative, about the present and future direction of the field. The comments fall under two headings, each describing two possible paths. The first heading rests on the observation that the past, present, and likely future direction of the field exemplifies a familiar distinction between two modes of innovation in science and technology. The second heading turns on a distinction between two possible paths for the science itself.

---

[27] Some studies report spectroscopic particle-energy signatures that appear unconventional, prompting additional theoretical and methodological questions.

[28] As an excellent recent example of the kind of experiment that might serve as a reference experiment, we note that of Ziehm (2022) and Ziehm and Miley (2024). The latter piece, released on the ArXiV preprint server in 2024, is also notable for another reason. For many years, ArXiV would not accept papers on cold fusion. The fact that it now does so is a mark of the change in the reputation of the field over the past decade.

## 7.1 Two modes of innovation

A distinction is often drawn between two different modes of innovation in science and technology, *Edisonian* and *rationalist* (Hughes 1977). This distinction is especially relevant where underlying physical mechanisms remain poorly understood or contested. Not surprisingly, therefore, cold fusion offers a good illustration.

The Edisonian mode involves iterative, trial-and-error experimentation, and design variation, in the absence of a well-worked-out theory of the domain in question. Famously, Edison tested thousands of different filament materials before arriving at a reliable incandescent light bulb. The solution was not deduced from a theoretical model, but found through systematic tinkering. This approach thrives in conditions of theoretical uncertainty or novelty, where the relevant physical variables are not yet fully known, and discovery proceeds by exploration rather than prediction.

In contrast, the rationalist mode of innovation presumes a well-established theoretical framework that guides design choices and optimization strategies. This is the paradigm of model-based engineering, where progress depends on reliable simulations, predictive equations, and clear cause-effect understanding. The development of modern microprocessors, commercial jet engines, or magnetic confinement fusion devices like tokamaks exemplifies this logic: designs are refined through theoretical models that have been validated by generations of experimental work.

In the cold fusion case, for example, Clean Planet[29] fits squarely within the Edisonian tradition. Faced with an anomalous phenomenon that defies easy theoretical explanation, researchers such as Iwamura and his colleagues have opted for high-precision experimental development, careful variation of materials and conditions, and long-term empirical pattern-seeking (see, e.g., Kitamura et al 2018). Their effort is not theory-free, but it is not theory-led in the traditional sense. Rather than deriving their designs from a settled physical model, they work inductively, seeking reproducibility and controllability first, and hoping that theory will follow.

In principle, the Edisonian approach may bypass ordinary gatekeeping in science. If it yields some sort of device that is undeniably real, and unexpected by the lights of existing theory, then it is up to theory to catch up. This has not yet happened in LENR; though it would happen, presumably, if Clean Planet or some other group released a working prototype reactor. In that case, the eventual maturation of the field would repeat the historical arc of other once-obscure technologies, beginning with empirical craft and only later acquiring theoretical foundations and rationally designed technology. At present, then, there are two paths by which the field might achieve 'normal' scientific status. Programs such as the ARPA-E's and their offshoots might succeed in defining an irrefutable reference experiment, or at least articulate a research agenda that is both satisfactory and sufficiently intriguing to contemporary gatekeepers. Or Clean Planet or some other group might release the equivalent of Edison's light bulb.

---

[29] And the post-2012 Japanese LENR research landscape more broadly. There are other, more recent examples elsewhere.

Reflecting on the history of the field, the Edisonian approach has been prominent. Apparent progress has come not from breakthroughs in fundamental understanding, but from painstaking refinement of experimental systems, often through decades of incremental, hands-on work. It also highlights one reason why the field has remained marginal in the eyes of mainstream science: it does not yet conform to the rationalist model of high-energy physics, say, or modern engineering.

This has been a stumbling block for the field. Too much of its focus has been consumed by trial-and-error tinkering towards small increases in excess thermal power. Too many researchers have pursued premature commercial ventures, hoping that the Edisonian method will yield a cold fusion boiler that detractors will be unable to ignore, hoping to bypass gatekeepers rather than engage with them. And too few have collaborated openly, within the field. As McKubre puts it in a 2017 survey:

> The understandable, but regrettably human, response to [the] opportunity [of LENR] has been in far too many instances the search for personal, regional or corporate glory, fame, wealth or power. This condition has petrified our progress by stultifying collaboration. I am no less responsible for – or less guilty than any other – of concealing secrets. The barbarians are not outside the gates – they are inside – they are us. (McKubre 2017, 15)

## 7.2 Two kinds of new science

What can be said about the possible future direction of the underlying science of LENR? We want to note three possibilities, and say something more about two of them. The first possibility is that there is no interesting new science at all. To the extent that they exist at all, claimed experimental anomalies can be attributed to systematic error, misinterpreted chemical processes, or uncontrolled experimental variables. This has been the dominant stance among critics since 1989, of course (to the extent that they have been paying attention at all). In our view, it has low probability at this point, but we won't argue the point here.

On the assumption that no prosaic explanation is possible, there are two main possibilities for the future direction of the science.

1. <u>Emergent behavior from known physical mechanisms</u> (Metzler et al. 2024) take this approach as their point of departure. Rather than invoking entirely new physics, the authors ask and review: What known mechanisms across atomic physics, nuclear physics, and quantum dynamics could plausibly take place and possibly coincide in condensed matter systems to enhance fusion rates?

In the atomic domain, mechanisms such as achieving temporarily greater deuteron proximity (Fork et al 2020) and electron screening effects (Czerski et al 2020) are well-established to influence fusion probabilities. These effects can contribute substantial rate enhancements, though not enough to account for most of the reported anomalies. In the nuclear domain, resonant reaction channels and compound nucleus states offer routes to further rate enhancement (Czerski et al 2024, Adsley et al 2022). And from a quantum optics perspective, coherent nuclear excitation transfer mechanisms – including Dicke-like models involving nuclear states (Terhune & Baldwin 1965, Bocklage 2021) – suggest that energy might be redistributed across coupled nuclei in a lattice in ways not captured by traditional two-body fusion approximations[30].

This line of inquiry remains well within the scope of known physics, and thus – per Ockham's razor – should be thoroughly pursued before resorting to more speculative alternatives.

2. <u>Signs of a paradigm shift</u>? Secondly, there remains the possibility that LENR phenomena signal a fundamental breakdown of current theoretical frameworks. Some proposed models invoke new particles, low-lying electronic states, or fundamentally modified interaction regimes. In our view, the threshold for adopting such interpretations should be high: they must be supported by reproducible experiments, be consistent with the wide range of well-established non-LENR experiments, and demonstrate explanatory power that clearly exceeds what the first category offers.

## 8. Conclusions – lessons for gatekeeping

In §§3–4 we criticised the gatekeeping of cold fusion in the early years. It was overly hasty, relied on a misleading notion of replication, and paid little heed to basic prudential principles ('Big Bang, Low Bar'). These mistakes are compounded by the derision encouraged by some influential actors, including *Nature.* This had the effects already documented by Goodstein in 1994, cutting the field off from funding, talent, and simple respect.

As we stressed at the beginning, these criticisms do not depend on hindsight. What can we add to them, with the benefits of hindsight? One is that as the field pursued its course in exile, it became increasingly probable that – setting haste, imprudence, and derision aside – the early gatekeeping simply got it wrong. It discarded some interesting science.

We don't mean that the case for *new* science was made with certainty. But twenty years ago, if not thirty years ago, there was already a nontrivial case for the existence of interesting anomalies, worth investigating by ordinary standards. In this very cautious form, this conclusion follows from the reactions of outsiders we described in §5 – some of the contributors to the 2004 DoE report, and Rob Duncan, for example. The relevance of the derision is that this was largely invisible to mainstream science at that point. Scientists of the time put their reputations at risk, if they proposed to take a look.

Events of the past decade have done nothing to weaken the case that the early gatekeeping made the wrong call (even, again, setting haste, imprudence, and derision aside). This recent period in the history of the field will be of lasting interest to students of gatekeeping for a different reason: it is an example of a field attempting to *recover* from gatekeeping mistakes. This part of the story isn't yet concluded, of course. As we saw in §7, there are several paths it might take. But what we already have on the table is fascinating enough – the involvement of Google, for example, and the careful subsequent role of ARPA-E.

---

[30] From that perspective, the continuous evolution of nuclear quantum optics may eventually encompass many reported LENR anomalies via models that readily predict modified nuclear reaction rates and modified reaction products in certain structured and stimulated environments such as those present in LENR experiments.

Price (2024a) says that if the field does recover in this way, as he expects, it will eventually be seen as exceptional in two respects. The first factor, he says, 'will be the depth of the reputational trap from which it will have managed to dig itself out.' (Here he includes: 'the severity of the condemnation and ridicule to which the field has been subject; its highly public nature; and the involvement of major scientific institutions such as *Nature* in administering it.') The second factor, in Price's view, is the prudential point. Few cases in the history of science come with such a high potential cost to a false negative.

We can summarise our assessment of what the gatekeeping of cold fusion got wrong in the following way – three apparent mistakes, with different levels of probability.

1. <u>With significant probability, difficult to assess accurately</u>: dismissing the source of an important technology, needed for a critical purpose. The uncertainty here gets in at several points. We don't yet know whether the pursued theoretical and experimental programs would lead to such applications, or how much difference they would make, if so, compared to alternatives.
2. <u>With high probability</u>: dismissing some interesting new science.
3. <u>With near-certainty</u>: a gross failure of the prudential principle, both in these dismissals themselves, and in the way in which it was done; overly hasty, poorly grounded on experimental evidence, and reckless in its abuse of reputation.

As we said at the beginning, the present account of the history of cold fusion is partial, in two senses. The history of the field is a work in progress, and we ourselves are not detached observers. Concerning the latter point, we noted that some readers, aware of the derided reputation of the field, might therefore be inclined to apply it to us. As we said, we welcome such reactions, as living manifestations of the history we have to present.[31] But we claim that such reactions are pathological, nevertheless – pathologies of the gatekeeping process, which would have been deplored and deprecated, if the case had been better handled.

**Acknowledgements:** We are grateful to the following for comments, discussion, and information: Geoff Anders, Melinda Baldwin, Philip Ball, Carlotta Barone-MacDonald, Melinda Bradley, Oliver Carefull, Rob Christian, Colin Clark, Harry Collins, Konrad Czerski, Matt DeCapua, Heather Douglas, Eman Elshaikh, Nicola Galvanetto, Peter Godfrey-Smith, Peter Hagelstein, Jirohta Kasagi, Ross Koningstein, Matt Lilley, Marianne Macy, Kim Milton, Carl Page, Alan Smith, and Matt Trevithick. We are also greatly indebted to the late Michael McKubre (1948–2025), one of the giants of the field, and a kind and generous man. We dedicate this piece to his memory.

**Disclosures:** The authors have no relevant financial interests to disclose. FM and JM are part of the MIT team under the US Department of Energy's ARPA-E LENR program. The authors' contributions are equal.

**Funding information:** Partial financial support was received from the Anthropocene Institute (JM, FM).

---

[31] We thank such readers for making it this far, and ask them to consider whether their own attitudes might not be products of gatekeeping mistakes, more than thirty years ago.